\documentclass[reprint,amsmath,amssymb,aps]{revtex4-1}
\usepackage{graphicx}
\usepackage{dcolumn}
\usepackage{bm}
\usepackage{natbib}
\usepackage[colorlinks,linkcolor=black,anchorcolor=blue,citecolor=blue,urlcolor=blue]{hyperref}

\begin{document}
\title{Surface relaxation of vapour-depositing colloidal glasses}
\author{Xin Cao, Huijun Zhang}
\author{Yilong Han}
\email{yilong@ust.hk}
\affiliation{Department of Physics, Hong Kong University of Science and Technology, Clear Water Bay, Hong Kong, China}
\date{\today}

\maketitle

\textbf{Vapour deposition can directly produce ultrastable glasses \cite{swallen2007science,singh2013nmat,dalal2015pnas}, which are similar to conventional glasses aged over thousands of years. The highly mobile surface layer is believed to accelerate the ageing process of vapour-deposited glasses, but its microscopic kinetics has not been experimentally observed. Here we studied the deposition growth kinetics of a two-dimensional colloidal glass at the single-particle level using video microscopy. We found that newly deposited particles in the surface layer (depth $d<14$ particles) relaxed via frequent out-of-cage motions, while particles in the deeper middle layer ($14<d\lesssim100$ particles) relaxed via activation of cooperative rearrangement regions (CRRs). These CRRs were much larger, more anisotropic and occurred more frequently than CRRs in the bulk ($d\gtrsim100$ particles) or after deposition. Their centers of mass moved towards the surface, while the particles within moved towards the bulk, causing free-volume ``bubbles'' to move towards the surface to give a more compact bulk glass. This two-step relaxation in two surface layers is distinct from the previously assumed relaxation in one surface mobile layer.} 

Conventional glasses formed from the quenching of liquids relax slowly to more stable structures via ageing. By contrast, vapour deposition can produce organic, polymeric and metallic glasses with extraordinary kinetic stability \cite{swallen2007science,singh2013nmat,dalal2015pnas}. Such ultrastable glasses can have highly uniform amorphous structures \cite{singh2013nmat}, unusually high \cite{swallen2007science,singh2013nmat} or low \cite{guo2012nmat} densities, enhanced elastic moduli \cite{kearns2010am,singh2013nmat,yu2013am} and highly anisotropic molecular orientations \cite{dalal2015pnas}. These properties are of significant interest in both practical material design and the theoretical understanding of the nature of glass transition. 
Experimental techniques such as differential scanning calorimetry \cite{swallen2007science}, neutron reflectivity \cite{swallen2007science}, dielectric measurements \cite{yu2015prl}, spectroscopic ellipsometry \cite{dalal2015pnas} and wide-angle X-ray scattering \cite{gujral2015cm} have been applied to study vapour-deposited glasses. 
These studies emphasized the importance of surface mobility in the formation of ultrastable glasses. 
A highly mobile, liquid-like surface layer exists on the free surface of organic \cite{zhu2011prl} and polymeric \cite{forrest1996prl, ellison2003nmat, priestley2005science, chai2014science} thin-film glasses. This surface mobile layer is responsible for the extremely low viscosity and glass transition temperature in polymer thin films \cite{ediger2013review}. It is conjectured that this mobile layer accelerates the rearrangement of newly deposited particles in a layer-by-layer fashion before they are buried into the bulk \cite{swallen2007science,kearns2007jcp,kearns2008jpcb},    
but the microscopic relaxation kinetics remains to be confirmed experimentally.

Colloids are outstanding model systems for the study of glasses because the real-space trajectories of individual particles can be measured by video microscopy \cite{hunter2012rpp,weeks2000science}. Colloidal glass studies have mainly focused on the bulk properties or confinement effects \cite{hunter2012rpp,hunter2014prl}, but have rarely explored free surfaces, i.e. vapour-glass interfaces. 
In this paper we study the relaxation near the free surface of a glass at the single-particle level for the first time by using colloids. In particular, we measured the kinetics during and after vapour deposition and found a mobile layer that was approximately 14 particles thick near the surface both during and after deposition. The newly deposited particles underwent frequent out-of-cage motion in the surface layer until they were buried into the bulk, which confirms the conjecture of a layer-by-layer relaxation in the deposition growth of glasses \cite{swallen2007science,kearns2007jcp,kearns2008jpcb}.
Interestingly, we discovered a middle layer ($14<d\lesssim100$ layers of particles) which relaxes via the emergence of many large cooperative rearrangement regions (CRRs). These CRRs propagated to the free surface, releasing free volumes to the vapour phase to give a more compact deposited glass. 

\begin{figure}[!bh]
  \centering
  \includegraphics[width=1\columnwidth]{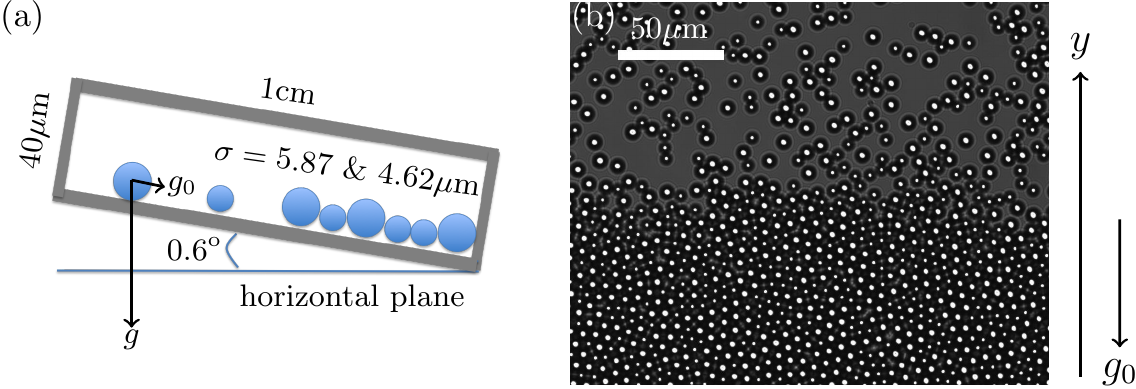}
  \caption{\textbf{A binary colloidal glass monolayer formed by vapour deposition (Movie 1).} (a) Schematic of the sample cell. (b) The colloidal glass grew along the y direction by vapour deposition under gravity. Scale bar: $50~\mu\textrm{m}$.}
  \label{fig1}
\end{figure}

\begin{figure*}
  \centering
  \includegraphics[width=1.8\columnwidth]{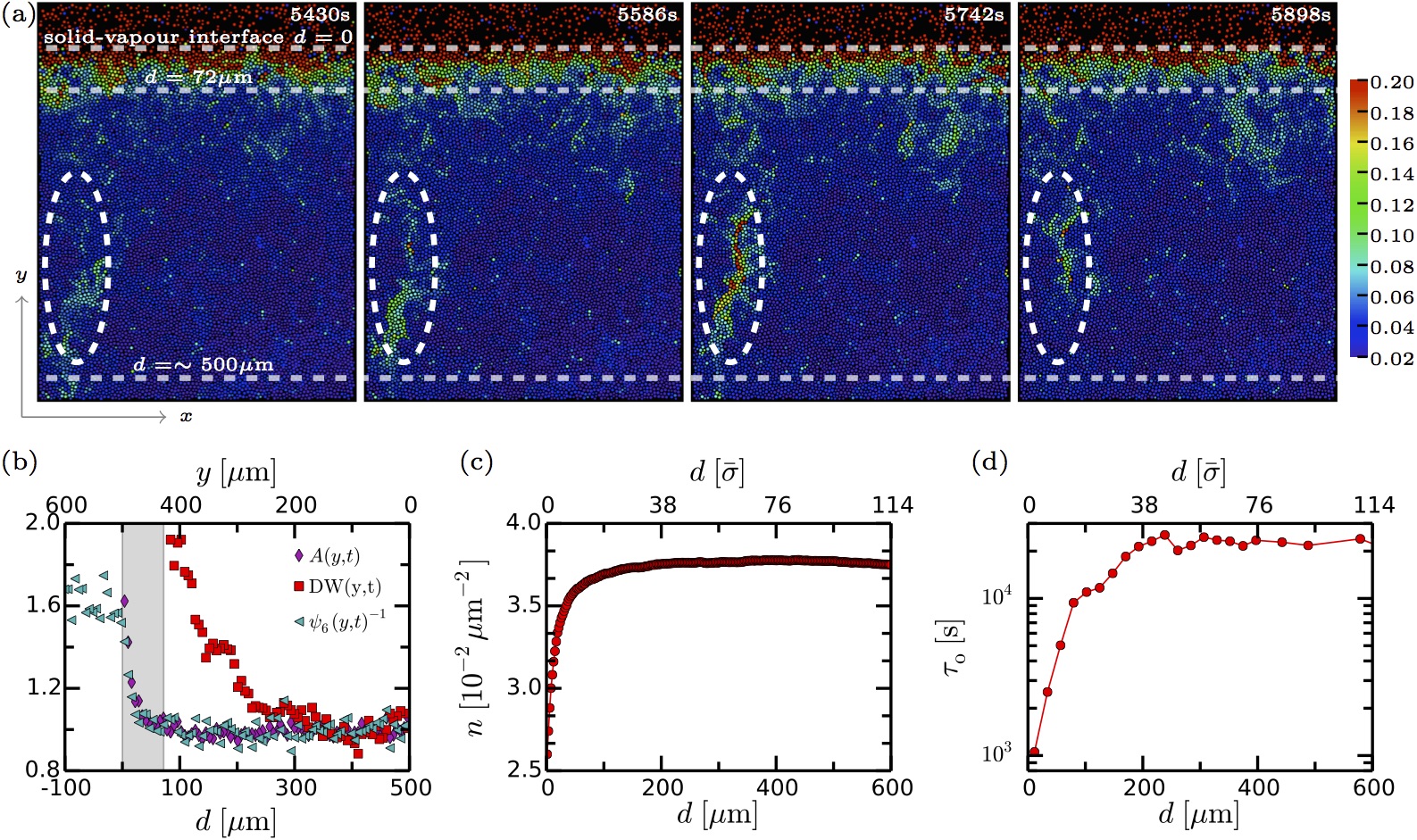}
  \caption{\textbf{Surface profiles of structural and dynamic properties during vapour deposition.} (a) Surface dynamics of the deposited monolayer glass from 5430~s to 5898~s during growth. Colours represent the value of the DW factor of each particle. $d=0$ marks the glass-vapour interface. $0<d<72~\mu\textrm{m}=14~\bar{\sigma}$ represents the highly mobile surface layer. In the middle layer of $72~\mu\textrm{m}<d\lesssim500~\mu\textrm{m}$, many clusters of large-DW-factor particles, i.e. CRRs, emerged. The dashed ellipses show the evolution of a CRR. The bulk at $d\gtrsim500~\mu\textrm{m}$ has similar DW values as the middle layer, but much rarer and smaller CRRs. (b) Average area spanned by Voronoi polygons $A(y,t)$, DW factor $\textrm{DW}(y,t)$ and inverse of the crystalline order $\psi_6(y,t)^{-1}$ normalized by the bulk values as a function of depth $d$. The difference between their profiles defines a surface mobile layer at $0<d<72~\mu\rm{m}$ marked by the gray region. At the boundaries of the gray region, $\textrm{DW}$ and $A$ are twice their corresponding bulk values. $\psi_6^{-1}$ follows a similar profile as $A$. (c) Profile of the particle density $n$. (d) Profile of the mean out-of-cage time. (c) and (d) were averaged over all frames.}
  \label{fig2}
\end{figure*}

We used a $45\%:55\%$ mixture of poly(methyl methacrylate) (PMMA) spheres with diameters $\sigma=4.62~\mu$m and  $5.87~\mu\rm{m}$. The  $2~\rm{cm}\times1~\rm{cm}\times40~\mu\rm{m}$ sample cell was slightly tilted at an angle of approximately $0.6^\circ$, as shown in Fig.~\ref{fig1}a. PMMA spheres with diameters $\sigma=4.62~\mu$m and $5.87~\mu$m and a mass density of 1.18~g/cm$^3$ have gravitational heights $k_\textrm{B}T/(mg)= 0.044~\mu$m and $0.021~\mu$m respectively, where $mg$ is the buoyant weight, $k_\textrm{B}$ is the Boltzmann constant and room temperature $T=295$~K. Consequently, they settled on the substrate with negligible motions along the $z$ direction and slowly drifted towards the lower end of the sample cell as shown in Fig.~\ref{fig1} and Movie 1. The clear-cut solid-vapour interface propagated slowly at a speed of $\nu_{y}=0.0112~\mu\rm{m/s}$ towards the vapour phase (Fig.~S1).

For each particle $i$ at time $t$, we characterized its local density by its inverse Voronoi area $1/A_i(t)$, the local sixfold crystalline order by $\psi_{6i}(t)$ and the strength of local dynamics by the Debye-Waller factor $\textrm{DW}_i(t)$ (see Supplementary Information (SI) for details). The local dynamics is shown in Fig.~\ref{fig2}a. Averaging these quantities over the $x$ direction yields their profiles along the $y$ direction as shown in Fig.~\ref{fig2}b and Movie 2. $A(y,t)$ and $\textrm{DW}(y,t)$ varied considerably at different depths, reflecting a surface mobile layer whose density is close to that of the bulk while the dynamics is much faster. The measured thickness of the surface mobile layer averaged over all frames is $72\pm 9~\mu\rm{m}=14\pm2~\bar{\sigma}$, where $\bar{\sigma}$ is the average diameter of spheres. This value is comparable to the atomic layer thickness of the surface mobile layer measured in a thin-film polystyrene glass at the glass transition temperature \cite{paeng2011jacs,ediger2013review}. 
 
The relaxation time of a glass is usually characterized by the decay time of the intermediate scattering function $F_s(q,t)$. We measured $F_s(q,t)$ from the long-time trajectories of particles in the bulk (Fig.~S5), but not near the surface because the surface was shifting during the deposition and different depths are associated with different relaxation rates. Instead we characterized the structural relaxation near the surface by the mean out-of-cage time $\tau_\textrm{o}$, i.e. the time interval between two consecutive out-of-cage events for a particle (see SI for details). The measured $\tau_\textrm{o}$ is much shorter in the surface layer (Fig.~\ref{fig2}d), reflecting more frequent out-of-cage events (Movie 1). Each particle stayed in the surface mobile layer for approximately $d_{\textrm{mob}}/\nu_y=6.4\times10^3$~s before it was buried into the middle layer, thus it experienced 5-10 out-of-cage events in the mobile layer since $\tau_{\rm{o}}\simeq 10^3$~s.
$\tau_{\rm{o}}$ is approximately $2.0\times10^4~\rm{s}$ in the middle layer and the bulk (Fig.~\ref{fig2}d), which agrees with the measured structural relaxation time $\tau_{\rm{s}}=1.4\times10^4$~s in the bulk (see Fig.~S5). 
When a particle is leaving the cage formed by its neighbours, the system is exploring the phase space from one local free-energy minimum to another. Therefore $1/\tau_{\rm{o}}$ is a measure of the rate of exploration in the phase space. 
During the deposition process, particles in the mobile layer quickly explore the phase space before they are buried into the bulk, forming an ultrastable glass layer-by-layer
\cite{swallen2007science,kearns2007jcp,kearns2008jpcb}. 
Fig.~\ref{fig2}d shows that the out-of-cage motion in the surface mobile layer is approximately 10 times faster than that in the middle layer, so the phase-space exploration in the former is much faster. Vapour-deposited glasses are less stable when the temperature is too low for a surface mobile layer to form, or when the deposition rate is too high for the surface mobile layer to relax fully \cite{swallen2007science, kearns2008jpcb}. Here we suggest that it is the thick slowly-relaxing middle layer rather than the thin surface mobile layer which determines the maximum deposition rate at which ultrastable glasses can form. If the middle layer is buried into the bulk before it is fully relaxed, the resulting glass will not be ultrastable.

\begin{figure}[!t]
  \centering
  \includegraphics[width=1\columnwidth]{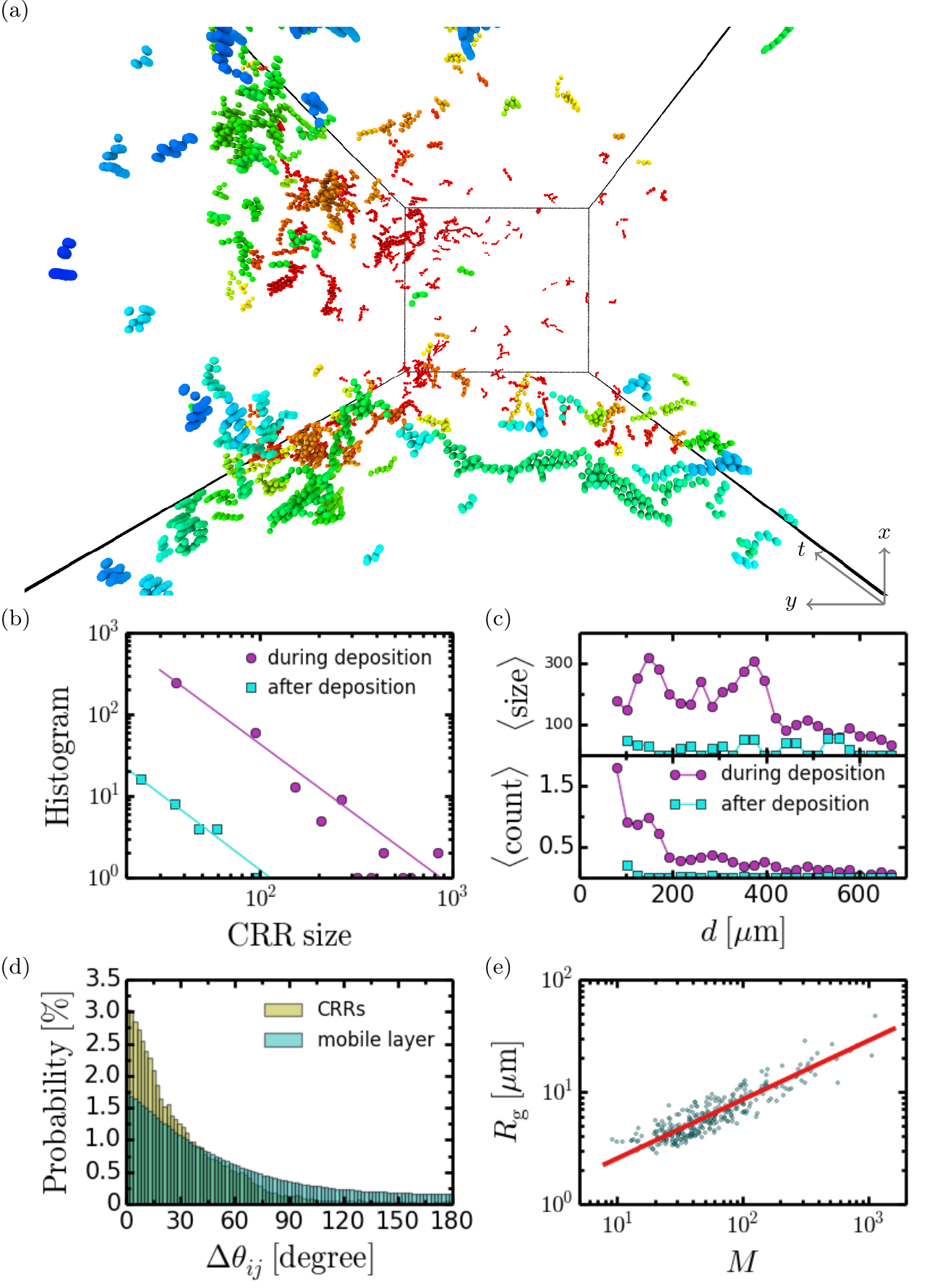}
  \caption{\textbf{CRR morphologies.} (a) CRRs in spacetime at $d>72~\mu\textrm{m}$ for a period of $\Delta t=34128$~s during deposition. Colours indicate time. The unit for space is $\bar{\sigma}$ and the unit for time is the characteristic time 18.1~s for a particle diffusing $\bar{\sigma}$ in the vapour phase. (b) Histograms showing the size of CRRs, i.e. the particle number in spacetime, fitted by power laws with exponents $\gamma=1.74\pm0.09$ and $\gamma=1.79\pm0.14$ during and after deposition respectively. (c) The average CRR size and CRR count are much higher at $72~\mu\textrm{m}<d\lesssim500~\mu\textrm{m}$ than in the deeper bulk, indicating that a middle layer existed during deposition but vanished afterwards. (d) Histograms showing the angle $\Delta\theta_{ij}$ between the displacements of particle $i$ and its neighbour $j$ during 156~s (Fig.~S3) in all the CRRs during deposition and in the surface mobile layer. (e) Radius of gyration $R_{\rm{g}}=\Sigma_i m_i(\vec{r}_i-\vec{r}_{\textrm{c}})^2/M$ of CRRs in spacetime during the deposition as a function of CRR mass $M=\Sigma_i m_i$, where $\vec{r}_{\textrm{c}}=\Sigma_i m_i\vec{r}_i/M$ is the position of the CRR's center of mass in spacetime, $m_i$ is the normalized area (i.e. $m_i=1$ for small particles) of particle $i$ and $\Sigma_i$ is the sum over all particles in spacetime.  The red line is the power-law fitting $R_{\rm{g}}\propto M^{1/d_{\textrm{c}}}$ with the fractal dimension $d_{\textrm{c}}=1.91\pm0.05$. CRR mass is different from CRR size because a larger sphere has more mass.}
  \label{fig3}
\end{figure}

\begin{figure*}
  \centering
  \includegraphics[width=1.8\columnwidth]{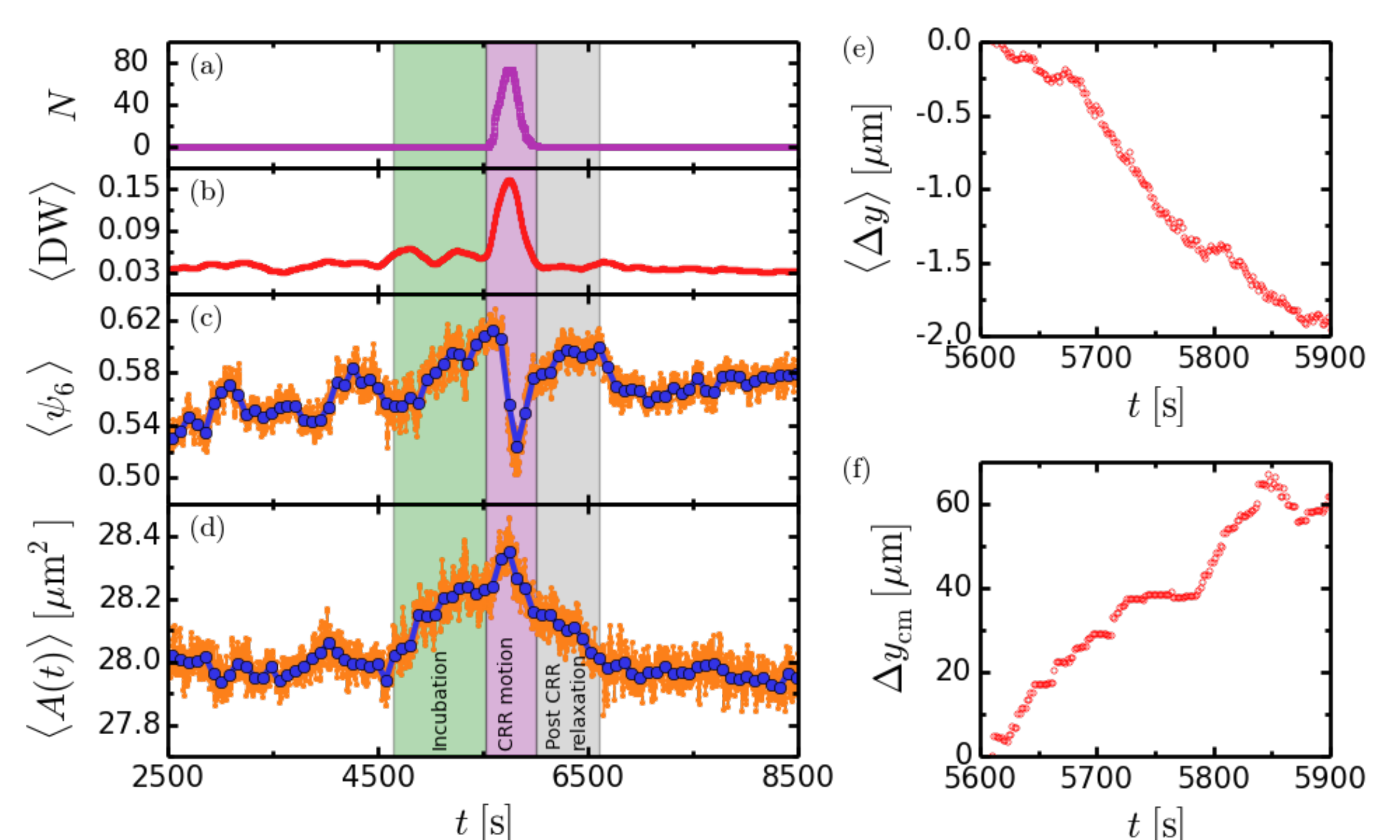}
  \caption{\textbf{Evolution of the CRR in Fig.~2a.} (a) Particle number $N$, (b) DW factor, (c) orientational order parameter $\langle \psi_6\rangle$ and (d) Voronoi cell area $\langle A \rangle$. $\langle\quad \rangle$ represents the average over all 80 particles in the CRR. The blue curves in (c, d) are smoothed over 50 bins of the raw data (orange curve).  (e) Average displacement $\langle\Delta y\rangle$ of the CRR particles as a function of $t$. $\Delta y$ is the displacement relative to the position at $5610$~s. (f) Displacement of the center of mass $\Delta y_{\rm{cm}}$ of the CRR relative to the position at $t=5610$~s.}
  \label{fig4}
\end{figure*}

As a key concept in glass relaxation, CRRs have been intensively studied in the bulk \cite{donati1998prl, hunter2012rpp}, but rarely near surfaces. Bulk CRRs are like strings at high temperatures or in systems composed of repulsive particles~\cite{stevenson2006nphy}, while form compact domains at low temperatures or in systems composed of attractive particles~\cite{stevenson2006nphy,zhang2011prl}. We measured the spacetime morphology of CRRs as shown in Fig.~\ref{fig3}a, which characterizes both the structure and dynamics of the local relaxations. $\textrm{DW}>0.12$ particles are defined as CRR particles, while different threshold values yield similar results (see the details in SI). Particles in the surface mobile layer moved rapidly and much less cooperatively, thus CRRs are not defined because otherwise the whole layer would be one huge CRR. The size of the CRRs in Fig.~\ref{fig3}a follows a power-law distribution as shown in Fig.~\ref{fig3}b. The power-law exponents in Fig.~\ref{fig3}b are close to the power-law exponent $5/3$ of the probability distribution of earthquake amplitudes, i.e. the Gutenberg-Richter law in seismology \cite{braun2013pre}. In fact, both CRRs and earthquakes are barrier crossing process involving the collective motions of densely packed materials.

We measured the density profile of the CRR count defined as the average number of CRR particles per frame at the instantaneous depth $d$ (lower panel of Fig.~\ref{fig3}c), i.e. the density profile of the colored particles along the y-axis in Fig.~\ref{fig3}a. Similarly, the CRR size profile defined as the average number of CRR particles at depth $d$ weighted by their CRR size is shown in the upper panel of Fig.~\ref{fig3}c.
During deposition, many more CRRs were emerged in the middle layer than in the bulk (Fig.~\ref{fig3}c), and they were also much larger in the middle layer, in agreement with the notion that the activation energy is lower near the surface \cite{stevenson2008jcp}. 
After deposition, CRRs in the middle layer became similar to bulk CRRs (Fig.~\ref{fig3}c), suggesting that the middle layer existed only during the deposition stage.

The collectiveness of particle motions in CRRs can be visualized in the wavy oscillation of DW factors in Movie 2 and can be characterized by the angle $\Delta\theta_{ij}$ between the displacements of neighboring particles $i$ and $j$. The histograms of $\Delta\theta_{ij}$ for CRR particles in the middle layer and in the surface mobile layer are shown in Fig.~\ref{fig3}d. $\Delta\theta_{ij}$ is distributed near 0$^\circ$ for CRR particles, indicating that neighbouring CRR particles tended to move in a similar direction, i.e. string-like motion \cite{weeks2002prl} as shown in Fig.~\ref{fig3}a. By contrast, $\Delta\theta_{ij}$ for fast particles in the mobile layer in Fig.~\ref{fig2}d show that their movements are much less cooperative. The CRRs in spacetime have fractal-like morphology with dimension $d_{\rm{c}}=1.91$ as shown in Fig.~\ref{fig3}e. Fractal CRRs have been predicted in mean-field theory \cite{klein1986prl, stevenson2006nphy} and observed in colloid experiments \cite{weeks2000science, zhang2011prl} and computer simulations \cite{donati1999pre, johnson1998pre} in space, whereas we found that CRRs are still fractals in spacetime.  
In addition, string-like CRRs tended to be perpendicular to the surface when they were moving towards the surface, e.g. the long axes of the ellipses in Fig.~\ref{fig2}a are perpendicular to the surface. This reflects the free-surface-induced symmetry breaking in both CRR structure and dynamics.  

The typical evolution of an individual CRR is shown in Fig.~\ref{fig4}. Figures~\ref{fig4}a-d reveal three stages of a CRR near the surface which have not been reported  in previous CRR studies which were in bulk \cite{stevenson2006nphy,zhang2011prl}. 
In the initial incubation stage, $\langle A \rangle$ increased by $0.2~\mu\textrm{m}^2$ and $\langle \psi_6 \rangle$ increased slowly, while $\langle\textrm{DW}\rangle$ maintained a low value but fluctuated more strongly before an 80-particle CRR (Fig.~4a) emerged. This 80-particle CRR region absorbed $\Delta A_{\textrm{total}}=0.2~\mu\textrm{m}^2\times80=16~\mu\textrm{m}^2$ of empty space from ambient regions. This amount of free volume is equivalent to 95\% of the area of a small particle or 59\% of the area of a large particle. This free volume triggered the collective motion, i.e. the CRR stage. The increased crystalline order (Fig.~\ref{fig4}c) made more efficiently use of the space which also helped to make room for the rearrangement. In the CRR stage, $\langle\textrm{DW}\rangle$ developed a sharp peak (Fig.~4b) with a similar shape to $N(t)$ (Fig.~4a). Meanwhile $\langle A \rangle$ rose to the maximum and $\langle \psi_6\rangle$ developed a sharp trough, indicating that the CRR was accompanied by a minimum local density and maximum local disorder. In the third stage, the dynamic quantity $\langle\textrm{DW}\rangle$ relaxed immediately back to the equilibrium value, corresponding to the vanishing of the CRR. However, $\langle A \rangle$ equilibrated slowly in 500~s. The observed three stages for a CRR reflect the evolution of structure and dynamics during a barrier-crossing process from one inherent structure to a more stable one. The three stages with their similar features have been observed for most of the large CRRs in the middle layer, but are difficult to resolve in bulk CRRs because of their small size. 
Note that the density decrease in the incubation stage is not necessary for the CRR to emerge as long as the free volume is large enough (see Fig.~S11 for more demonstrations).

Interestingly, we found that particles in a CRR tended to drift towards the bulk, while the center of mass of the CRR propagated towards the surface. These novel behaviors are shown in Figs.~\ref{fig4}e, f for an individual CRR and in Fig.~S10 for the ensemble average over all CRRs. A CRR can be viewed as a bubble of free volume since it is less dense than the ambient region (Fig.~\ref{fig4}d). We observed that a free surface could attract bubbles within a depth $d\lesssim100~\bar{\sigma}$ and caused these bubbles to float towards the surface. Eventually the bubbles were released into the vapour phase via CRRs, rendering a more compact bulk glass. By contrast, bulk CRRs propagated randomly and could not change the mesoscopic density of the glass. 

Small free volumes between particles in the middle layer can hardly migrate towards the free surface individually. However, when they nucleated and triggered a CRR motion, their propagation towards the surface was dramatically enhanced. The observed free-volume nucleation agrees with the free-volume picture of viscous liquid which described that molecular transport occurs via the movement of molecules into voids when the voids are larger than some critical size \cite{cohen1959}. 

The two relaxation mechanisms during deposition consist of the early-stage particle diffusion in the surface mobile layer and the later-stage release of free volumes towards the surface via CRR in the middle layer. Since a free surface should naturally induce these two effects, we expect that they occur commonly in various vapour-deposited glasses. Note that free-volume bubbles were released only during deposition. After deposition, the middle-layer regime became indistinguishable from the bulk, which indicates a well-equilibrated glass after the bubble-releasing relaxation. These behaviors were confirmed in our simulation of bidispersed disks with Weeks-Chandler-Andersen potential: a CRR-rich middle layer $14 \lesssim d \lesssim 100~\bar{\sigma}$ formed below the $d\lesssim 14~\bar{\sigma}$ diffusive surface mobile layer during deposition but disappeared quickly afterwards (Movie 3) (see SI for details of the simulation). These results cast new light on the formation of ultrastable glasses and the effect of surfaces on glass relaxation.

\textbf{Acknowledgement:}
The work was supported by RGC grants GRF601613, A-HKUST616/14 and C6004-14G-A. We thank Shenshen Wang for useful discussions.

\onecolumngrid
\newpage

\large\textbf{Supporting information of ``Surface relaxation of vapour-depositing colloidal glasses''}

\makeatletter 
\renewcommand{\thefigure}{S\@arabic\c@figure}
\makeatother
\setcounter{figure}{0}

\makeatletter 
\renewcommand{\theequation}{S\@arabic\c@equation}
\makeatother
\setcounter{equation}{0}

\section{Experiment}
\begin{figure}[!h]
  \centering
  \includegraphics[width=0.5\columnwidth]{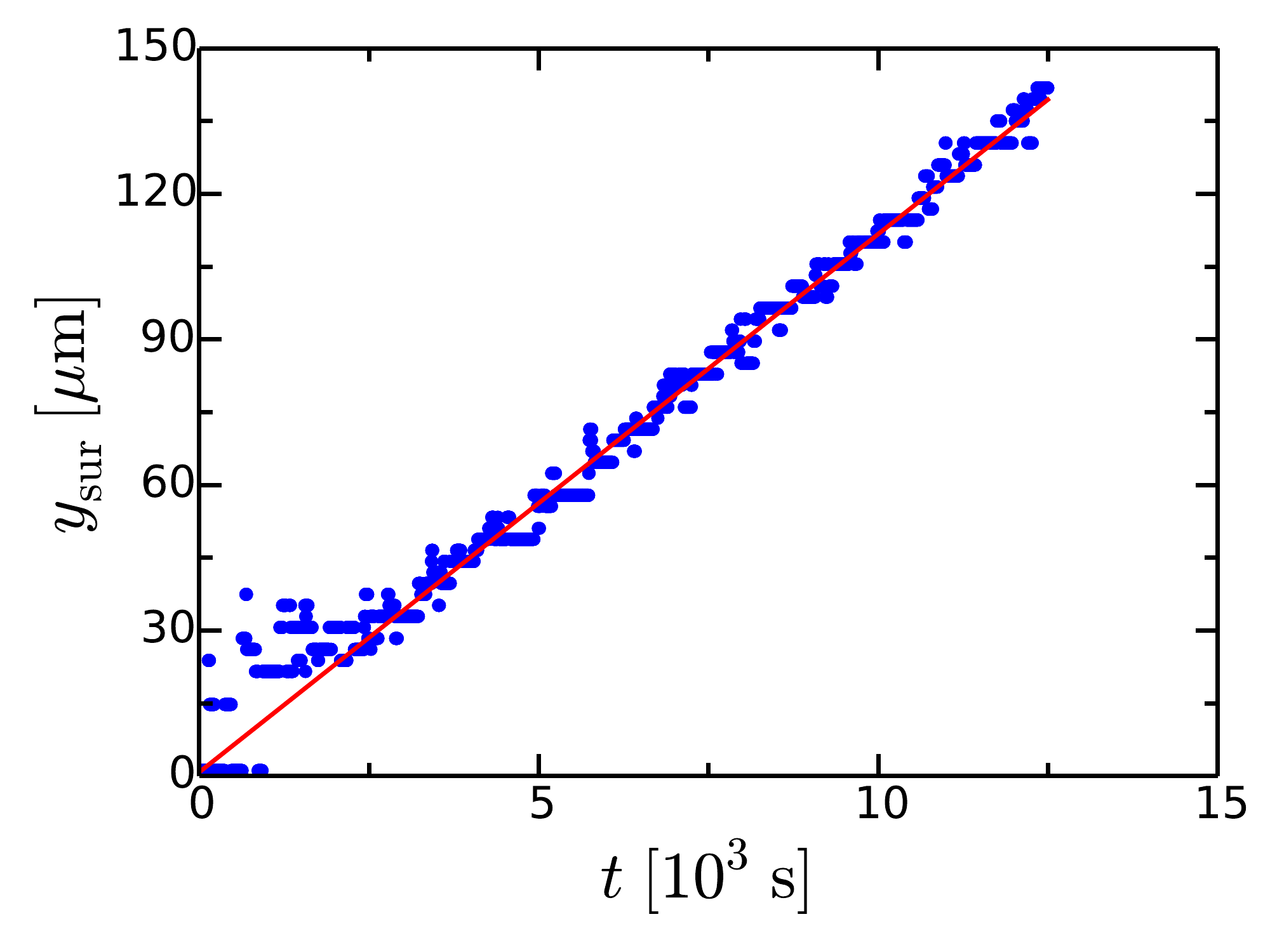}
  \caption{Position of the surface $y_{\rm{sur}}$ as a function of time $t$ during vapour deposition. The slope of the linear fit (red line) yields the propagation speed $\nu=0.0112~\mu\rm{m/s}$.}
  \label{fig:front_t}
\end{figure}

We mixed poly(methyl methacrylate) (PMMA) spheres with the diameters of $4.62\pm0.11~\mu\textrm{m}$ ($\it{microParticles\ GmbH}$ PMMA-R-B1170) and  $5.87\pm0.14~\mu\rm{m}$ ($\it{microParticles\ GmbH}$ PMMA-R-B1298) at a ratio of 0.45 to 0.55. Two pieces of $40~\mu\rm{m}$-thick paraffin films were sandwiched between a glass slide and a glass coverslip as spacers, forming a $20~\rm{mm}\times10~\rm{mm}\times40~\mu\rm{m}$ channel. A $10~\mu\rm{L}$ drop of colloidal suspension was placed at one end of the channel and pulled into the channel by the capillary force. The whole sample was sealed with an epoxy adhesive. Colloidal solution in a properly sealed sample can be kept for a few months without drying. The whole microscope including the sample cell on the microscope stage was placed on a rotational rack and tilted at an angle of $0.6^\circ$ with respect to the horizontal. At such a small tilt angle, the vapour phase of PMMA particles with a uniform area fraction of $\phi=23.8\%$ slowly settled towards the lower end of the sample cell (Movie 1 and Fig.~1b) and formed a monolayer glass with a clear-cut glass-vapour interface. The interface propagated toward the vapour side at a constant speed of $\nu=0.0112~\mu\rm{m/s}$ over a prolonged period as shown in Fig.~\ref{fig:front_t}.

\begin{figure}
  \centering
  \includegraphics[width=0.5\columnwidth]{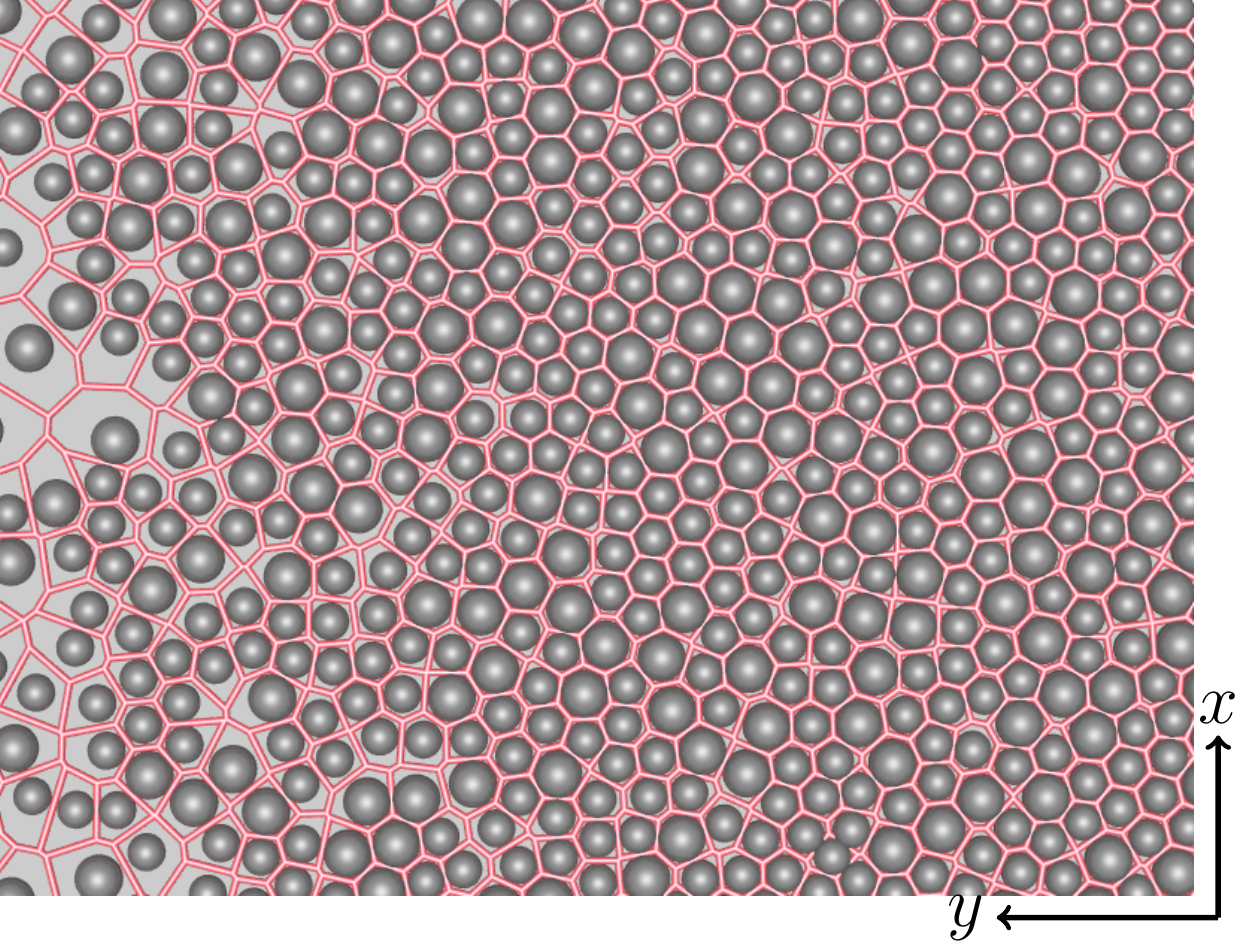}
  \caption{Radical Voronoi tessellation for the bidispersed monolayer near the glass-vapour interface.}
  \label{fig:radical}
\end{figure}

\section{Radical tessellation, Voronoi area and local crystalline order}
The traditional Voronoi tessellation can characterize the local density of a monodisperse sphere system, but not the local density of a dense binary system because the bisecting plane between two neighbouring spheres may cut through a large sphere and cannot reflect the real cage associated with each particle. Consequently we used the radical Voronoi tessellation in the Voro++ library \cite{rycroft2009voro++}. In the radical Voronoi tessellation, the radical plane is composed of the points with the same tangent length for the two neighbouring spheres, i.e. the tangential line from the point to each sphere has the same length. Radical Voronoi tessellation avoids the intersection with spheres and keeps the main topological features of the traditional Voronoi tessellation. It has been well adapted in grain hindrance and segregation problems \cite{gervois2002iccs}. An example of the radical Voronoi tessellation of our 2D binary system is shown in Fig.~\ref{fig:radical}. The inverse area $1/A_j$ and the number of edges $N_{nj}$ of the radical Voronoi polygon of particle $j$ give the local density and the number of neighouring spheres. 
The profile of the mean Voronoi area along the $y$ direction shown in Fig.~2b is calculated as $A(y,t)=\int A(x,y,t)\textrm{d}x /\int \textrm{d} x$.
The local crystalline order $\psi_{6j}=\Sigma_{k=1}^{N_{nj}}e^{-i\cdot6\theta_{jk}}/6$, where $N_{nj}$ is the number of neighbours of particle $j$, $\theta_{jk}$ is the angle of the bond between particle $j$ and its neighbour $k$ and $i^2=-1$.
Similarly, the $\psi_6$ profile along the $y$ direction in Fig.~2b is calculated as $\psi_6(y,t)=\int \psi_6(x,y,t)\textrm{d}x/\int \textrm{d} x$.

\section{Debye-Waller (DW) factor}
\begin{figure}[!h]
  \centering
  \includegraphics[width=0.5\columnwidth]{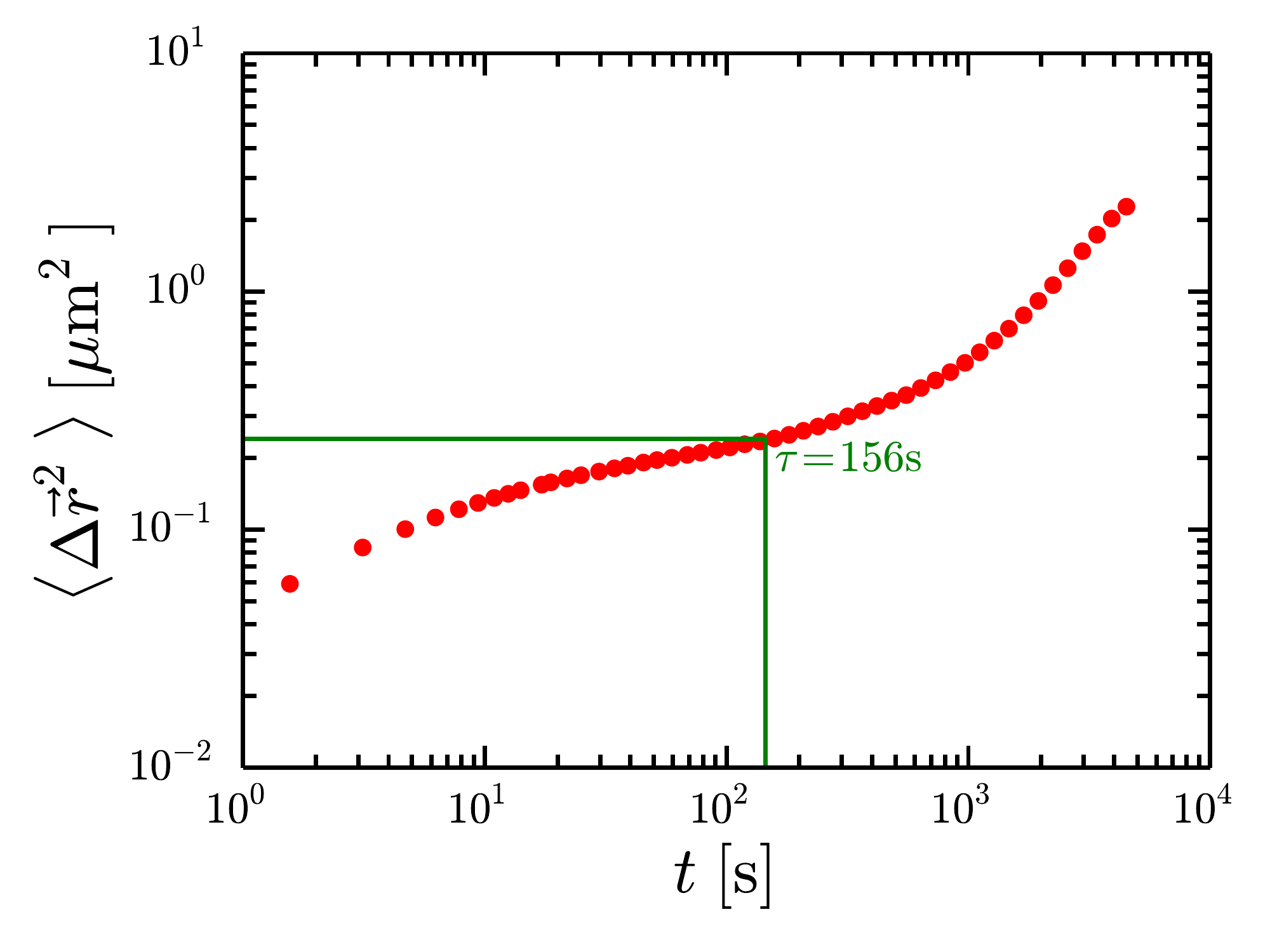}
  \caption{Mean-square displacement averaged over approximately 800 particles in the bulk. $\tau=156$~s at the plateau is used to calculate the DW factor.}
  \label{fig:msd_bulk}
\end{figure}

\begin{figure}
  \centering
\includegraphics[width=0.8\columnwidth]{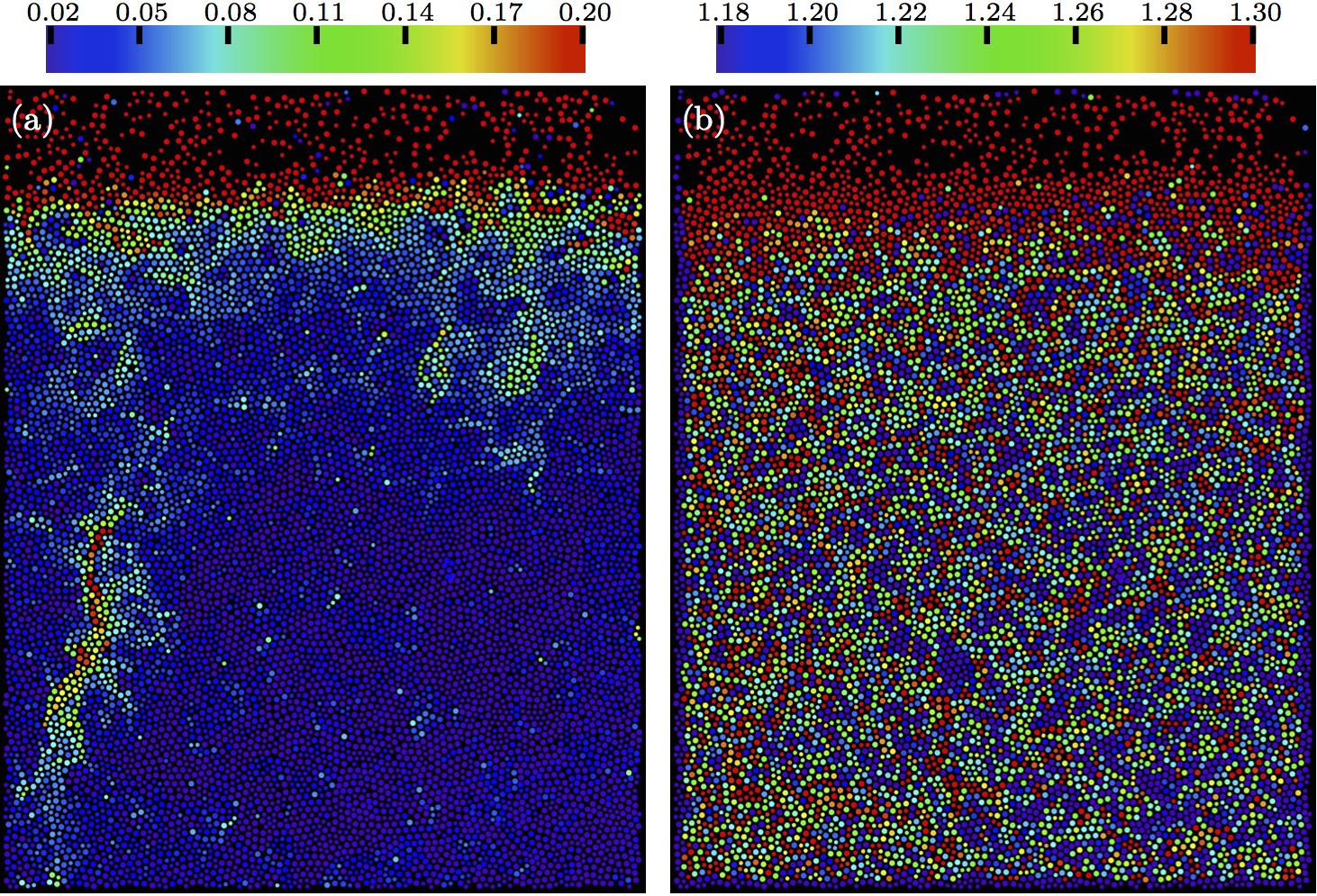}
  \caption{The same sample image coloured according to the DW factor (a) and the Voronoi area (b). The Voronoi areas of large and small particles are normalized by their corresponding particle areas. The surface mobile layer can be seen in both versions where the dynamics is fast in (a) and the density is low in (b). However the middle layer featuring large anisotropic CRRs can only be seen in (a) because CRRs have much stronger dynamics but only slightly lower densities than the ambient regions.}
  \label{fig:DWvoro}
\end{figure}

The dynamics of particle $j$ can be characterized by the modified Debye-Waller factor \cite{larini2007nphy} defined as $\textrm{DW}_j(t)= 2\sqrt{\langle\textrm{MSD}\rangle_{\tau}}/(\sigma_1+\sigma_2)=2\sqrt{\langle(\overrightarrow{x}_j - \langle\overrightarrow{x}_j\rangle_{\tau})^2\rangle_{\tau}}/(\sigma_1+\sigma_2)$, where $\langle \quad\rangle_{\tau}$ denotes the average over a time period $(t-\tau,t+\tau)$. $\tau=156$~s is the time needed for the mean-square displacement (MSD) of bulk particles to reach the plateau of the curve (see Fig.~\ref{fig:msd_bulk}). This choice of $\tau$ can clearly distinguish uncaged particles with a DW$_j\ge0.12$.
The profile of the mean DW factor along the $y$ direction is shown in Fig.~2a. Figure~\ref{fig:DWvoro} shows that the DW factor can reveal the CRRs and hence the middle layer, but the local density cannot.

\section{Out-of-cage time}
\begin{figure}
  \centering  \includegraphics[width=0.4\columnwidth]{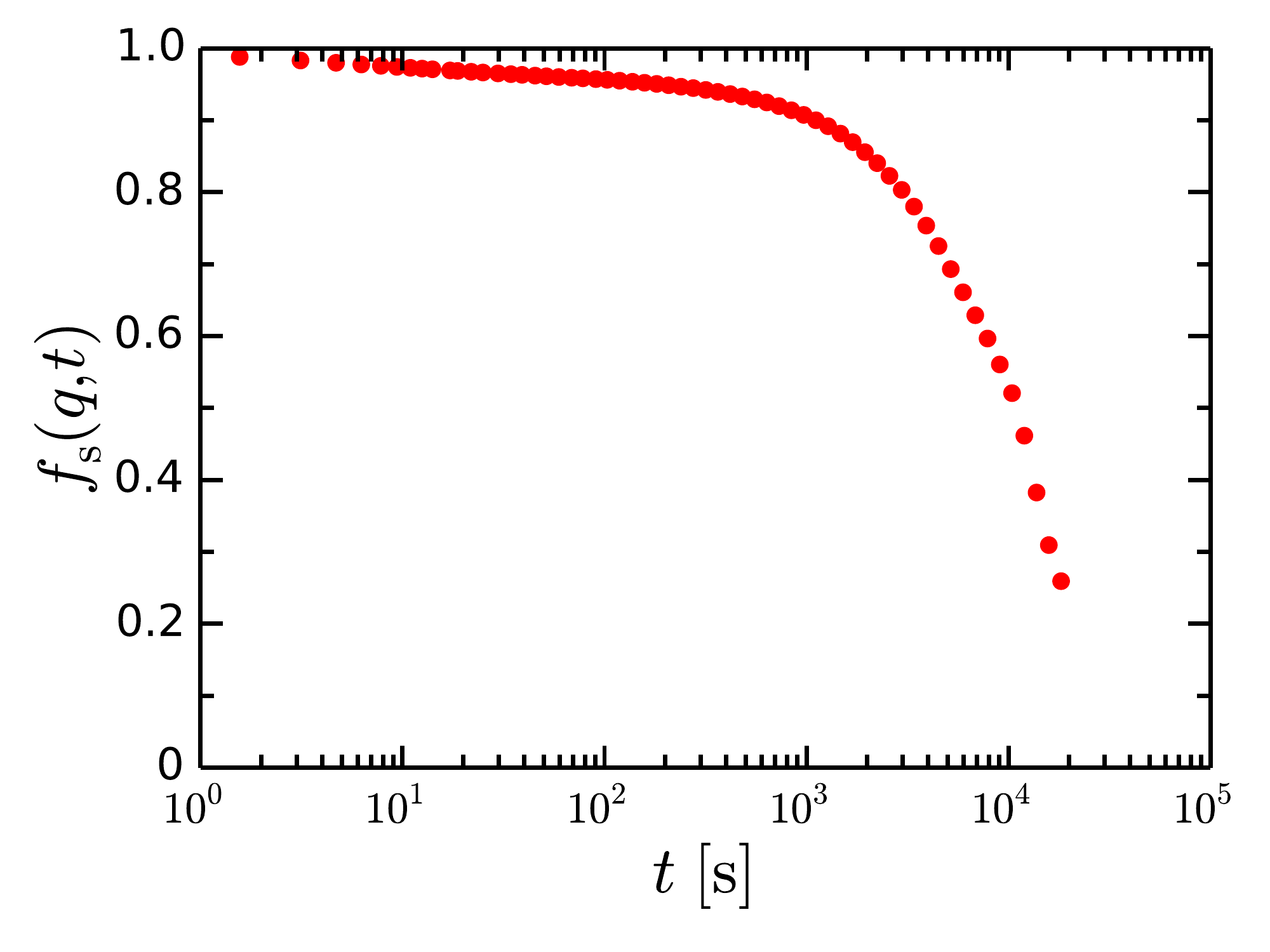}
  \caption{The self-part of the intermediate scattering function $f_{\textrm{s}}(q,t) $ in the $d\gtrsim500~\mu\rm{m}$ bulk. $q=4\pi/(\sigma_1+\sigma_2)$. The relaxation time $\tau_{\textrm{s}}=1.4\times10^4~\textrm{s}$ is defined as the time when $f_{\textrm{s}}(q,t)$ decays to $0.4$.}
  \label{fig:fsqt_bulk}
\end{figure}

\begin{figure}
  \centering \includegraphics[width=0.4\columnwidth]{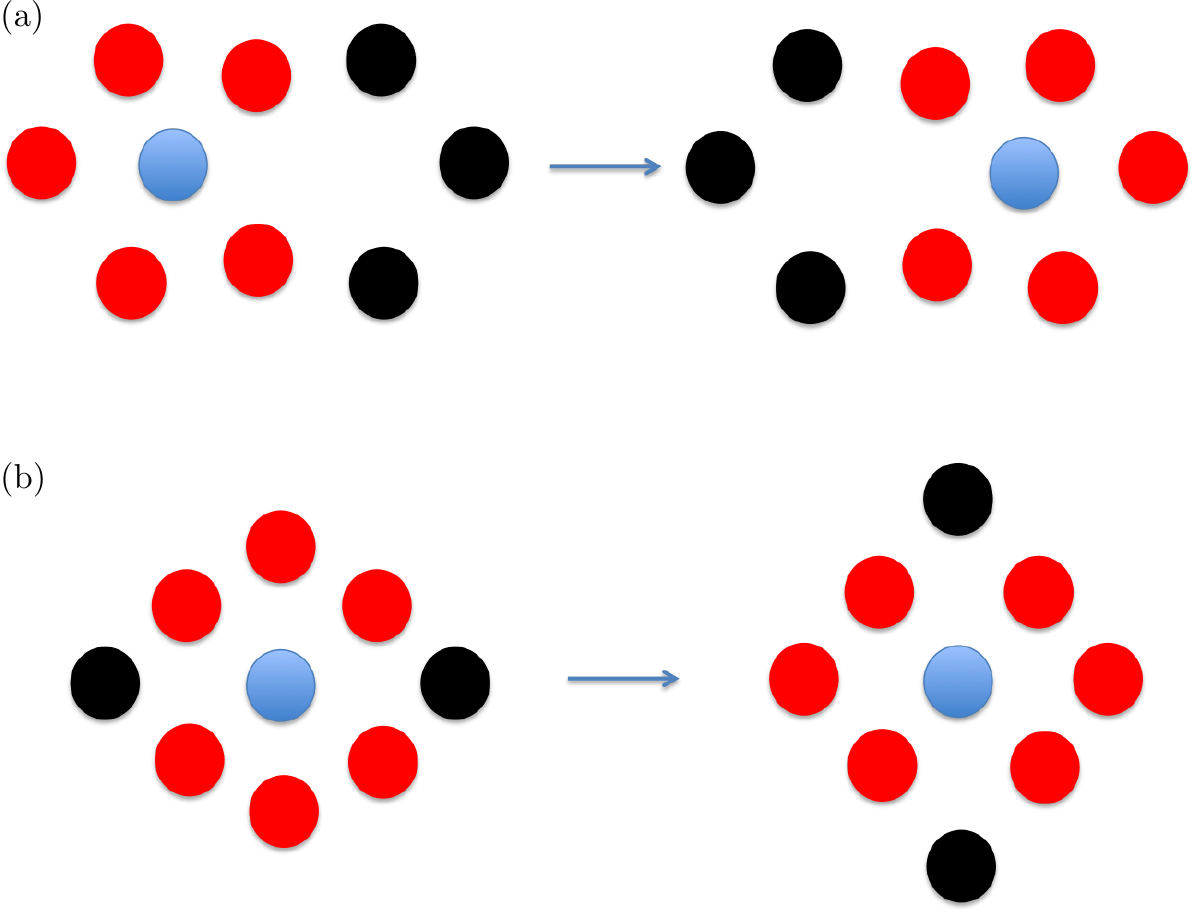}
  \caption{(a) When a particle leaves its original cage, three of its neighbours are replaced. Red particles represent the nearest neighbours. (b) A cage dilation replaces two neighbours of a particle, but the particle remains in the original cage. }
  \label{fig:out-of-cage}
\end{figure}

The self-part of the intermediate scattering function $f_{\textrm{s}}(q,t)=\langle\Sigma_{j=1}^N e^{i\vec{q}\cdot[\vec{r}_j(t)-\vec{r}_j(0)]}\rangle/N$ characterizes the structural relaxation in glassy systems \cite{kob1995}. $f_{\textrm{s}}(q,t)$ averaged over time is inappropriate when the surface is moving because it would mix different relaxation times at different depths. Therefore $f_{\textrm{s}}(q,t)$ can only be measured properly in the deep bulk or after deposition where the relaxation time stays constant (see Fig.~\ref{fig:fsqt_bulk}
for an example). Alternatively we define an out-of-cage time $\tau_{\rm{o}}$ to characterize the structural relaxation rate during deposition. $\tau_{\rm{o}}$ is defined as the time interval between two out-of-cage events, i.e. three of the nearest neighbouring particles will have been replaced since the last out-of-cage event. Here we set the threshold to three neighbours because a particle has six neighbours on average in 2D and an out-of-cage event will replace at least half of those neighbours (Fig.~\ref{fig:out-of-cage}a). The modification of two neighbours only may reflect a cage dilation instead of an out-of-cage event as shown in Fig.~\ref{fig:out-of-cage}b. 

\section{Characterization of the cooperative rearrangement region (CRR)}

\begin{figure}
  \centering
  \includegraphics[width=0.5\columnwidth]{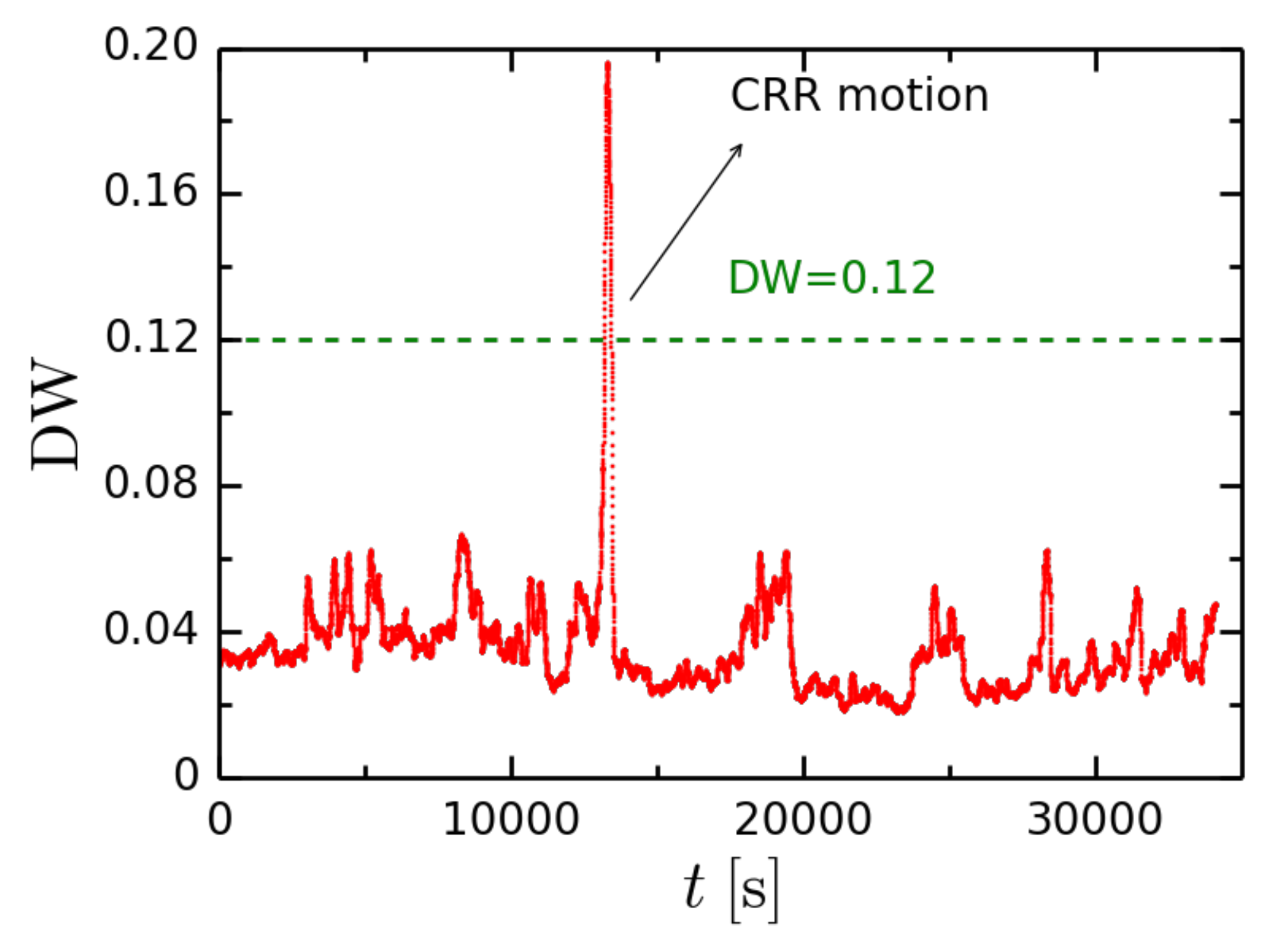}
  \caption{Typical evolution of the DW factor of a particle. The spike indicates that the particle was undergoing CRR.}
  \label{fig:lt}
\end{figure}

\begin{figure}
  \centering
\includegraphics[width=0.6\columnwidth]{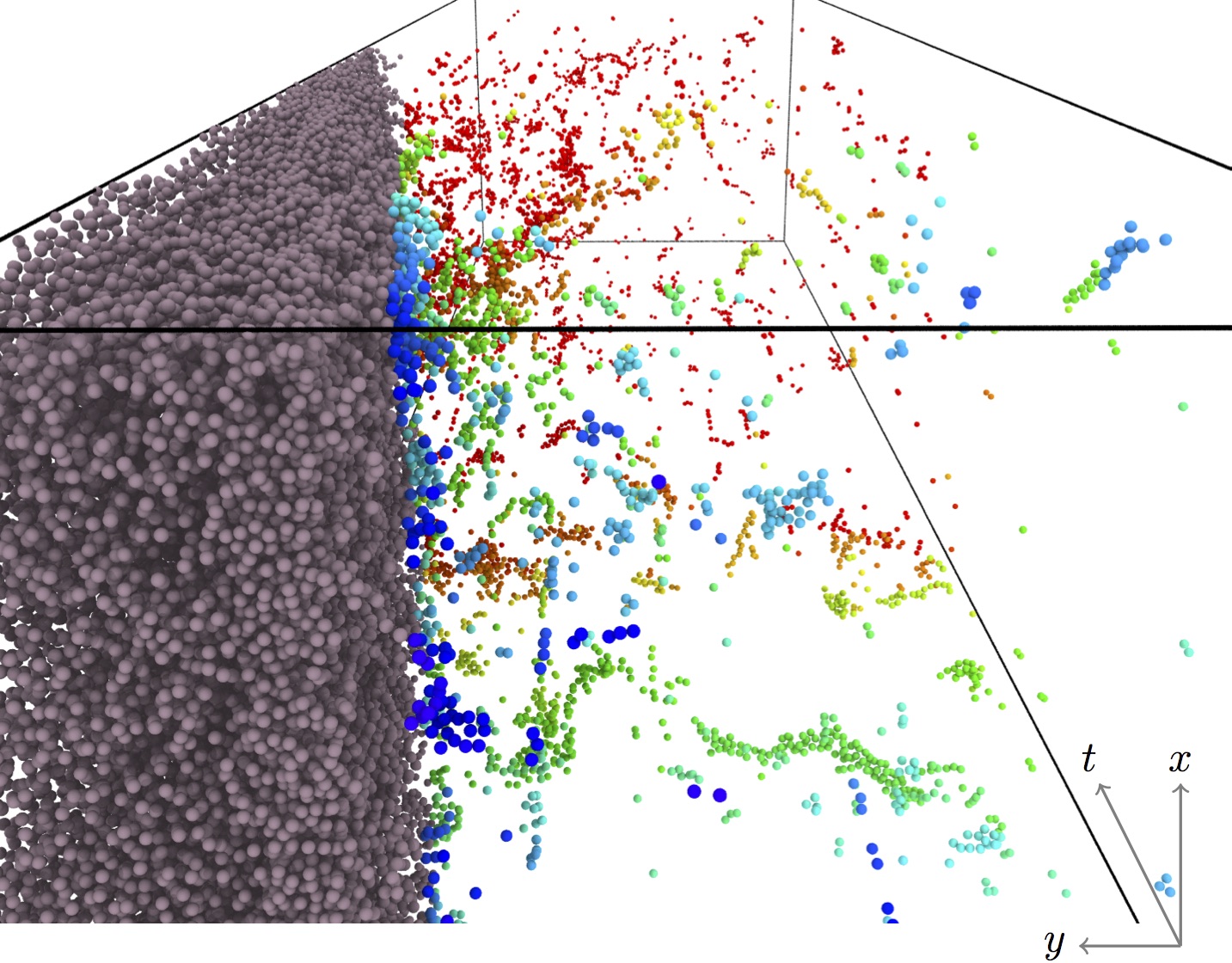}
  \caption{Particles with $\textrm{DW}\ge0.12$ are shown in spacetime. The colours indicate time order. These particles tended to form clusters, i.e. CRRs, suggesting that cooperative rearrangement is the dominant way of relaxation under the surface mobile layer. Gray particles are in the surface mobile layer or in the vapour phase, where CRRs are not well defined. Fig.~3a in the main text only shows the  middle layer and the bulk region without the gray particles.}
  \label{fig:spacetime}
\end{figure}

\begin{figure}
  \centering
  \includegraphics[width=1.0\columnwidth]{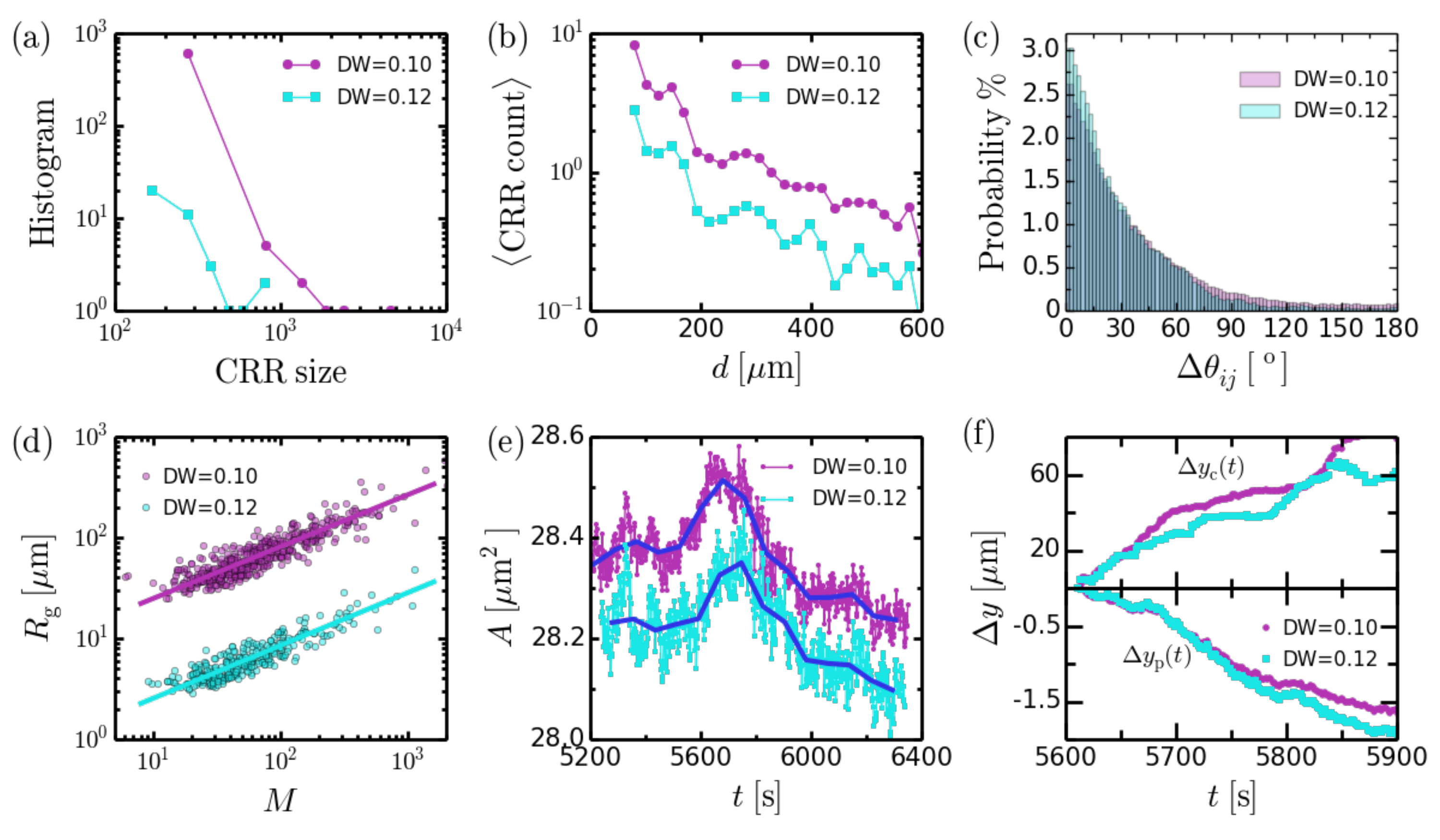}
  \caption{CRR particles identified from two different thresholds of the DW value, 0.10 and 0.12, exhibit similar CRR properties. (a) Histograms of the CRR size. (b) Average number of CRRs as a function of depth $d$. (c) Probability distribution of the angle difference between a particle's displacement and its neighbour's displacement during a period of 156~s. (d) Radius of gyration $R_{\rm{g}}$ as a function of CRR mass. The power-law fittings (red lines) give the fractal dimensions of 1.95 and 1.91 for DW values of 0.10 and 0.12 respectively. $R_{\rm{g}}$ for a DW value of 0.10 is multiplied by 10 to avoid overlap. (e) Time evolution of the average Voronoi area $A$ for the particles in the CRR shown in Fig.~2a. (f) The center of mass of the CRR in Fig.~2a propagated toward the surface, i.e. the displacement $\Delta y_{\textrm{c}}>0$ (upper panel), while the CRR particles moved into the bulk, i.e. $\Delta y_{\textrm{p}}<0$ (lower panel).}
  \label{fig:compare}
\end{figure}

When a particle is undergoing cooperative rearrangement, its DW factor will increase significantly (see the peak in Fig.~\ref{fig:lt}). In the middle layer and the bulk, 0.112\% of particles have a $\textrm{DW}>0.12$ and 0.282\% of particles have a $\textrm{DW}>0.10$. We define a mobile particle as one whose $\textrm{DW}>0.12$. These mobile particles formed CRRs below the surface mobile layer as shown in Fig.~S7. The critical value of approximately 0.12 is in accordance with the Lindemann criterion that a crystal melts when the mean amplitude of particle vibrations exceeds approximately 10\% of the lattice constant \cite{zheng1998epl}. We define CRRs as clusters of mobile particles in spacetime which can better reflect their time evolution (see Fig.~\ref{fig:spacetime}). By contrast, the conventional CRR defined as a cluster of mobile particles in space \cite{weeks2000science,zhang2011prl} is a cross-section of the CRR in spacetime. The time period for a DW factor exceeding 0.12 at the peak (e.g. Fig.~\ref{fig:lt}) is defined as the CRR time of a particle. Two particles are involved in the same CRR if they are neighbours in space and their CRR times overlap. We only consider CRRs involving more than one particle. The threshold of $\textrm{DW}=0.12$ or 0.10 in the definition of mobile particles yields similar results about CRRs as shown in Fig.~\ref{fig:compare}. 

\begin{figure}
  \centering
  \includegraphics[width=0.5\columnwidth]{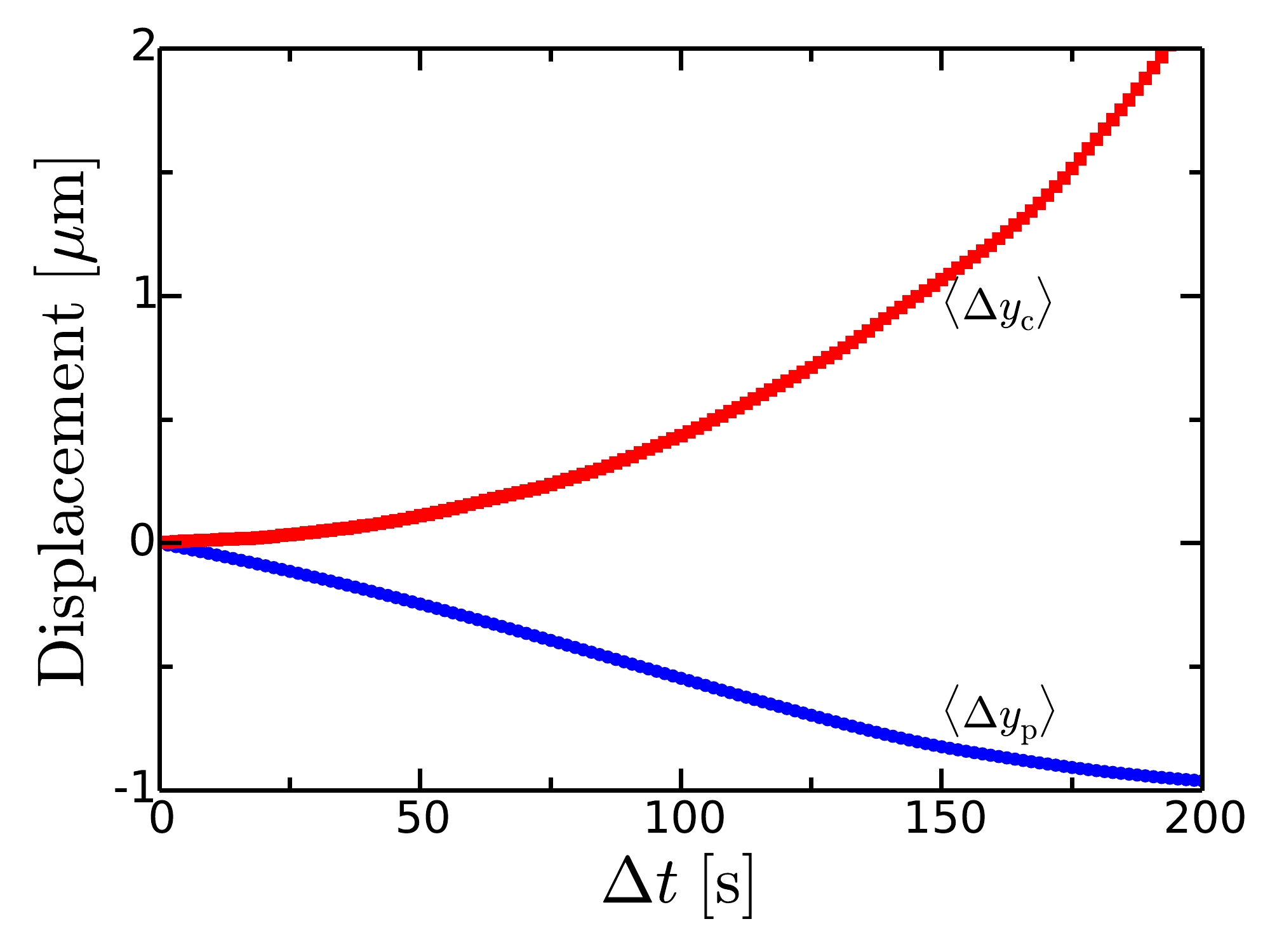}
  \caption{Displacements of CRR centers of mass $\langle\Delta y_{\textrm{c}}\rangle$ and CRR particles $\langle\Delta y_{\textrm{p}}\rangle$ averaged over all CRRs in spacetime. CRR centers of mass propagated toward the surface ($\langle\Delta y_{\textrm{c}}\rangle>0$), while CRR particles moved into the bulk ($\langle\Delta y_{\textrm{p}}\rangle<0$).}
  \label{fig:dydt}
\end{figure}

Movie 1 shows that CRRs tend to propagate toward the surface while CRR particles tend to move into the bulk. We quantified their motions for a single CRR in Fig.~4b, c and for the ensemble-averaged displacements over all CRRs in Fig.~\ref{fig:dydt}.

\begin{figure}
  \centering
 \includegraphics[width=0.8\columnwidth]{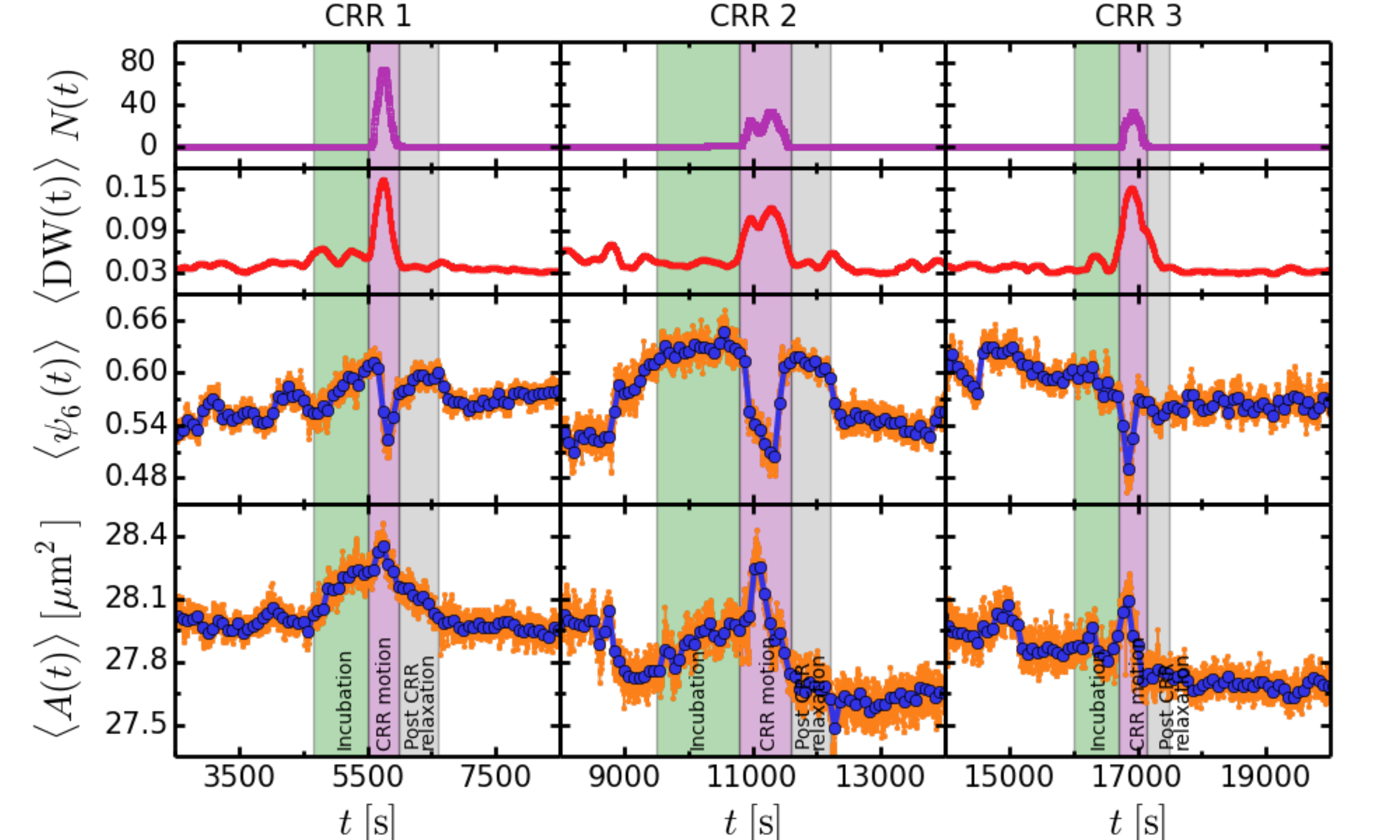}
  \caption{Time evolution of the number of CRR particles $N$, the DW factor, the crystalline order $\langle \psi_6 \rangle$ and the average Voronoi area $\langle A \rangle$ as a function of time for the three largest CRRs. $\langle\quad \rangle$ denotes the average over all particles in the CRR. The results suggest that CRRs are free-space bubbles (lower $A$ than the surroundings) with a more disordered structure (lower $\psi_6$). During a CRR event, a free-space bubble is released toward the surface, effectively reducing the local density.}
  \label{fig:bubbleSI}
\end{figure}

\section{Computer simulation}
We performed Brownian dynamics simulation of the deposition processes shown in Movie 3. A CRR-rich middle layer approximately 100 atomic layers thick emerged during deposition but disappeared afterwards, which agrees with our experimental observations. The simulation system consisted of a 50\%:50\% binary mixture of particles interacting with Weeks-Chandler-Andersen (WCA) potential $U(r)=4\epsilon[(\sigma/r)^12-(\sigma/r)^6+1/4]$ \cite{weeks1971jcp} with $m_{\textrm{A}}=1.0$, $m_{\textrm{B}}=2.0$, $\sigma_{\textrm{AA}}=1.0$, $\sigma_{\textrm{BB}}=1.3$, $\sigma_{\textrm{AB}}=1.15$, and $\epsilon_{\textrm{AA}}=\epsilon_{\textrm{AB}}=\epsilon_{\textrm{BB}}=\epsilon=100$. The simulations were conducted in $NAT$ ensemble (constant number of particles $N$, area $A=l_x \times l_y$, and temperature $T$) with $N=15000$ particles in a box measuring $l_x=130$ in width and $l_y=650$ in length. The periodic boundary condition was applied in the $x$ direction. Gravity was along the $y$ direction. The ground at $y=0$ was a WCA wall and the end of the vapour side at $y=650$ was a reflection wall. A time step of $\textrm{d}t = 0.0025$ was used in all simulations. The temperature was kept constant using a Nose-Hoover thermostat. After the vapour was fully equilibrated at $k_{\textrm{B}}T=0.02\epsilon$ in the absence of gravity, the temperature was decreased to $k_{\textrm{B}}T=0.005\epsilon$, and the gravitational weights $G_{\textrm{A}}=0.0025\epsilon/\sigma_{\textrm{AA}}$ and $G_{\textrm{B}}=0.005\epsilon/\sigma_{\textrm{AA}}$ were turned on along the $-y$ direction. The deposition process lasted for about $10^7$ steps until all vapour particles had been deposited on the glass.

\section{videos}
Movie 1: 
The deposition process corresponding to Fig.~2a. In the surface mobile layer, liquid-like particles frequently moved out of their cages. Under the mobile layer was a middle activation layer which contained large string-like CRRs. These CRRs propagated towards the free surface and released free volumes toward the surface.\\

Movie 2: 
Evolution of the profiles of the Voronoi area $A(d,t)$ and the DW factor $\textrm{DW}(d,t)$. The field of view was fixed during deposition, thus the depth $d$ changed as the free surface propagated. At a small $d$, the mobile layer has a similar density to the bulk but its DW factors are much higher. The DW factor in the middle activation layer shows strong and correlated fluctuations, representing the collective motions of CRRs. The surface propagated outside the field of view at the end.\\

Movie 3:
The deposition process in the simulation. A surface mobile layer comprising approximately 10-20 particle layers was present during and after deposition. A CRR-rich middle layer comprising approximately 100 particle layers was present during deposition but quickly disappeared afterwards.

\end{document}